\documentclass[twocolumn]{apex}

\usepackage{amsmath,amsthm,amssymb,mathrsfs}
\usepackage{graphicx} \usepackage{bm} \usepackage{amsmath}
\usepackage{amssymb}

\title{Magnetization switching by microwaves initially rotating in opposite direction to precession}

\author{
  Tomohiro Taniguchi
} 
\inst{
 National Institute of Advanced Industrial Science and Technology (AIST), Spintronics Research Center, Tsukuba, Ibaraki 305-8568, Japan 
}

\abst{
A common understanding of magnetization switching in 
microwave-assisted magnetization reversal is that 
the rotation direction of the microwaves should be the same with the precession direction of the magnetization. 
In this letter however, we show that microwaves initially rotating opposite to the magnetization precession 
destabilize the magnetization at an equilibrium 
and induces the switching more efficiently, 
when the microwave frequency depends on time. 
This argument is analytically deduced from energy balance equation. 
We also establish a model satisfying this condition, 
and confirm magnetization switching solely by microwaves by using numerical simulation. 
}
\begin{document}

\maketitle


When a rotating magnetic field originating from microwaves is applied to a ferromagnet, 
and the microwave frequency satisfies certain conditions, 
the ferromagnet efficiently absorbs energy from the microwaves, 
and the magnetization switches its direction from one stable state to the other. 
This phenomenon is called microwave-assisted magnetization reversal 
and has attracted much attention 
because of its advantageous writing scheme in magnetic recording media 
\cite{bertotti01,bertotti01a,thirion03,denisov06,sun06,nozaki07,zhu08,igarashi09,wang09,bertotti09,bertotti09book,okamoto08,okamoto10,okamoto12,tanaka13}. 
Microwave-assisted magnetization reversal has been confirmed 
by both experiments and numerical simulations. 


A common understanding of microwave-assisted magnetization reversal is that 
the rotating direction of the microwaves should be same as 
the precession direction of the magnetization. 
Note that the precession direction of the magnetization is determined by 
the field torque, $-\gamma \mathbf{m} \times \mathbf{H}$, acting on the magnetization, 
where $\gamma$, $\mathbf{m}$, and $\mathbf{H}$ are 
the gyromagnetic ratio, the unit vector along the magnetization direction, and the magnetic field, respectively. 
The microwaves rotating in the same direction as this precession direction efficiently supply energy to the ferromagnet 
and increase the precession amplitude. 
On the other hand, switching cannot be achieved by microwaves rotating 
opposite to the precession direction. 
This fact guarantees selective switching 
in microwave-assisted magnetization reversal \cite{suto14}. 


In this letter, we study the magnetization dynamics excited by microwaves 
initially rotating opposite to the precession direction. 
Based on the energy balance equation, 
we analytically show that such microwaves can induce an unstale condition for the magnetization in equilibrium 
when the microwave frequency is time dependent. 
We propose a model of the system having a time-dependent microwave frequency to satisfy this analytical prediction. 
In the model, the microwaves initially rotate opposite to the precession direction for a certain period. 
After the initial magnetization state is destabilized, 
the microwaves change their rotating direction, and synchronize with the magnetization precession. 
Numerical simulation of the Landau-Lifshitz-Gilbert (LLG) equation 
confirms magnetization switching solely by such microwaves. 



\begin{figure}
\centerline{\includegraphics[width=0.8\columnwidth]{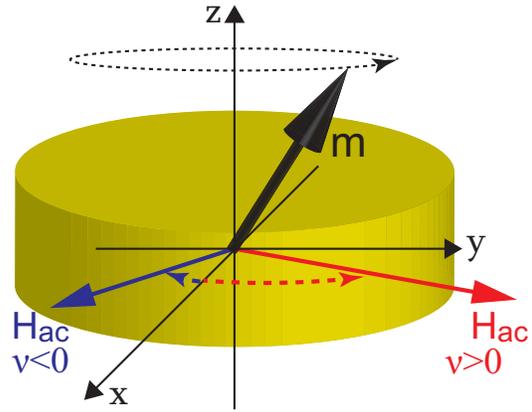}}
\caption{
         Schematic view of the system under consideration. 
         The magnetization $\mathbf{m}$ precesses around the $z$ axis, 
         where the precession direction is determined by the precession torque, $-\gamma \mathbf{m} \times \mathbf{H}$. 
         \vspace{-3ex}}
\label{fig:fig1}
\end{figure}



Figure \ref{fig:fig1} schematically shows the system under consideration. 
The magnetization dynamics in a ferromagnet are described by the LLG equation given by 
\begin{equation}
  \frac{d \mathbf{m}}{dt}
  =
  -\gamma
  \mathbf{m}
  \times
  \mathbf{H}
  -
  \alpha
  \gamma
  \mathbf{m}
  \times
  \left(
    \mathbf{m}
    \times
    \mathbf{H}
  \right),
  \label{eq:LLG}
\end{equation}
where $\alpha$ is the Gilbert damping constant. 
We assume that $\alpha$ is sufficiently small, i.e., $1+\alpha^{2}\simeq 1$. 
Let us focus on the switching of a uniaxially magnetized ferromagnet. 
In this case, the magnetic field $\mathbf{H}$ consists of 
a uniaxial anisotropy field along the easy ($z$) axis $H_{\rm K}$ 
and a microwave field rotating in the $xy$-plane as 
\begin{equation}
  \mathbf{H}
  =
  H_{\rm ac}
  \cos
  \psi
  \mathbf{e}_{x}
  +
  H_{\rm ac}
  \sin\psi
  \mathbf{e}_{y}
  +
  H_{\rm K}
  m_{z}
  \mathbf{e}_{z},
  \label{eq:field}
\end{equation}
where $H_{\rm ac}$ and $\psi$ are 
the amplitude and phase of the microwave field, respectively. 
The microwave frequency is defined as $\nu \equiv (d \psi /dt)/(2\pi)$. 
In the absence of microwaves, 
the ferromagnet has two stable states at $\mathbf{m}=\pm\mathbf{e}_{z}$. 
The precession direction of the magnetization due to the field torque, 
i.e., the first term on the right-hand side of Eq. (\ref{eq:LLG}), 
is a counterclockwise (clockwise) rotation with respect to the $z$ axis 
when the initial magnetization starts from the positive (negative) $z$-direction. 
The rotation direction of the microwave field is 
counterclockwise (clockwise) for a positive (negative) frequency $\nu$; see Fig. \ref{fig:fig1}. 


For further discussion, it is convenient to use a rotating frame $x^{\prime}y^{\prime}z^{\prime}$, 
in which the $z^{\prime}$-axis is parallel to the $z$-axis 
and the $x^{\prime}$-axis is always in the same direction as the microwave field \cite{bertotti09book}. 
The LLG equation in the rotating frame is given by 
\begin{equation}
\begin{split}
  \frac{d \mathbf{m}^{\prime}}{dt}
  =&
  -\gamma
  \mathbf{m}^{\prime}
  \times
  \bm{\mathcal{B}}
  -
  \alpha
  \gamma
  \mathbf{m}^{\prime}
  \times
  \left(
    \mathbf{m}^{\prime}
    \times
    \bm{\mathcal{B}}
  \right)
\\
  &+
  \alpha
  \frac{d \psi}{dt}
  \mathbf{m}^{\prime}
  \times
  \left(
    \mathbf{e}_{z^{\prime}}
    \times
    \mathbf{m}^{\prime}
  \right),
  \label{eq:LLG_rotating_frame}
\end{split}
\end{equation}
where the unit vector along the magnetization in the rotating frame is denoted as 
$\mathbf{m}^{\prime}=(m_{x^{\prime}},m_{y^{\prime}},m_{z^{\prime}})$. 
The magnetic field in the rotating frame is 
\begin{equation}
  \bm{\mathcal{B}}
  =
  H_{\rm ac}
  \mathbf{e}_{x^{\prime}}
  +
  \left(
    -\frac{1}{\gamma}
    \frac{d\psi}{dt}
    +
    H_{\rm K}
    m_{z^{\prime}}
  \right)
  \mathbf{e}_{z^{\prime}}.
  \label{eq:field_rotating_frame}
\end{equation}
As pointed out by Ref. \cite{taniguchi14}, 
Eq. (\ref{eq:LLG_rotating_frame}) has a mathematical structure that is analogous to 
the LLG equation with a spin torque \cite{slonczewski96,berger96}, 
where the last term of Eq. (\ref{eq:LLG_rotating_frame}) corresponds to the spin torque term 
with a pinned layer pointing in the $z$-direction. 
Therefore, let us, for the moment, call the last term in Eq. (\ref{eq:LLG_rotating_frame}) the spin torque for convention. 
This spin torque term moves the magnetization to the positive (negative) $z$-direction 
when the frequency is positive (negative). 
This means that the spin torque corresponding to 
the microwaves rotating opposite to the precession direction 
acts as an antidamping spin torque 
and tries to move the magnetization from equilibrium, $\mathbf{m}=\pm\mathbf{e}_{z}$. 
The phenomenon is due to the fact that such an antidamping spin torque does not favor a steady precession of the magnetization. 
However, the microwaves rotating opposite to the precession direction do not usually result in magnetization switching. 
This is because such microwaves at the same time makes the switched state energetically unstable, 
and the amplitude of the magnetization from equilibrium becomes relatively small \cite{taniguchi14}. 


We revisit this point by considering a time-dependent frequency. 
Note that, while experiments in past studies used microwaves with a constant frequency, 
the possibility is emerging to realize a time-dependent frequency 
by using an arbitrary wave generator or a spin torque oscillator (STO) coupled to the target ferromagnet \cite{suto14}. 
Several theoretical models have been proposed to show magnetization switching 
by microwaves with a time-dependent frequency \cite{rivkin06,barros11,barros13,klughertz15,taniguchi15APEX}. 
The previous theories are based on a resonant switching model \cite{rivkin06}, 
the variation method \cite{barros11,barros13}, 
and the autoresonance model \cite{klughertz15}. 
We recently proposed another type of resonant switching, 
in which the microwave frequency is always slightly different from the precession frequency \cite{taniguchi15APEX}. 
Macrospin and/or micromagnetic simulations confirmed switching of the magnetization 
by microwaves with a time-dependent frequency 
\cite{rivkin06,barros11,barros13,klughertz15,taniguchi15APEX,kudo15}. 
These works, however, assume that 
the microwaves always rotate in the same direction as those of the magnetization precession. 
Below, let us consider the magnetization dynamics by microwaves 
rotating opposite to the precession direction with a time-dependent frequency, 
based on an extended model of our previous work \cite{taniguchi15APEX}. 


To investigate the possibility of exciting the magnetization dynamics, 
it is convenient to study energy change of the system from the LLG equation \cite{taniguchi15APEX}. 
The energy density in the rotating frame is defined as $\mathscr{E}=-M \int d \mathbf{m}^{\prime}\cdot\bm{\mathcal{B}}$, 
where $M$ is the saturation magnetization. 
Then, from Eq. (\ref{eq:LLG_rotating_frame}), 
the energy change, 
$d \mathscr{E}/dt=(d \mathbf{m}^{\prime}/dt)\cdot(\partial \mathscr{E}/\partial \mathbf{m}^{\prime}) + (\partial \mathscr{E}/\partial t)$, is described as 
\begin{equation}
\begin{split}
  \frac{1}{\gamma M}
  \frac{d \mathscr{E}}{d t}
  =&
  -\alpha
  \left(
    -\frac{1}{\gamma}
    \frac{d \psi}{d t}
    +
    H_{\rm K}
    m_{z^{\prime}}
  \right)
  H_{\rm K}
  m_{z^{\prime}}
  -
  \alpha 
  H_{\rm ac}^{2}
\\
  &\ \ \ +
  \alpha 
  \left[
    H_{\rm ac}
    m_{x^{\prime}}
    +
    \left(
      -\frac{1}{\gamma}
      \frac{d \psi}{d t}
      +
      H_{\rm K}
      m_{z^{\prime}}
    \right)
    m_{z^{\prime}}
  \right]
\\
  &
  \ \ \ \ \ \ \ \ 
  \times
  \left(
    H_{\rm ac}
    m_{x^{\prime}}
    +
    H_{\rm K}
    m_{z^{\prime}}^{2}
  \right)
\\
  &\ \ \ 
  +
  \frac{1}{\gamma^{2}}
  \left(
    \frac{\partial}{\partial t}
    \frac{d \psi}{dt}
  \right)
  m_{z^{\prime}} .
  \label{eq:dEdt}
\end{split}
\end{equation}
The magnitude of the microwave field is usually much smaller than the uniaxial anisotropy field. 
Therefore, the dominant part of the energy change near the initial state, $\mathbf{m}\simeq \pm \mathbf{e}_{z}$, is 
\begin{equation}
\begin{split}
  \frac{1}{\gamma M}
  \frac{d \mathscr{E}}{dt}
  \sim&
  -\alpha
  H_{\rm K}^{2}
  m_{z^{\prime}}^{2}
  \left(
    1
    -
    m_{z^{\prime}}^{2}
  \right)
\\
  &+
  \alpha
  H_{\rm K}
  \frac{1}{\gamma}
  \frac{d \psi}{dt}
  m_{z^{\prime}}
  \left(
    1
    -
    m_{z^{\prime}}^{2}
  \right)
\\
  &+
  \frac{1}{\gamma^{2}}
  \left(
    \frac{\partial}{\partial t}
    \frac{d \psi}{dt}
  \right)
  m_{z^{\prime}}.
  \label{eq:dEdt_initial_main}
\end{split}
\end{equation}
Note that $d \mathscr{E}/dt$ should be positive to switch the magnetization. 
The first term is always negative because it comes from the damping torque. 
The second and third terms appear because of the presence of the microwaves. 
When the microwaves rotate in the same (opposite) direction as the precession direction, 
the second term is positive (negative) 
because $(d \psi/dt)m_{z^{\prime}} \propto \nu m_{z^{\prime}}>0$. 
Therefore, this term contributes to the increase (decrease) in the energy 
when the microwaves rotate in the same (opposite) direction as the magnetization precession. 
This is the common understanding in microwave-assisted magnetization reversal. 
On the other hand, the third term is finite when the microwave frequency is time dependent. 
This term can be positive even when the microwaves rotate opposite to the precession direction. 
An example can be found by considering the phase 
\begin{equation}
  \psi
  =
  c 
  \gamma 
  H_{\rm K}
  \left(
    m_{z}
    +
    \epsilon
  \right)
  t, 
  \label{eq:phase}
\end{equation}
where $c$ and $\epsilon$ ($\in \mathbb{R}$) are assumed to be constant. 
The physical meanings of these parameters are discussed below. 
The microwave frequency is given by 
\begin{equation}
  \nu
  =
  c 
  \frac{\gamma}{2\pi}
  H_{\rm K}
  \left(
    m_{z}
    +
    \epsilon
    +
    \frac{d m_{z}}{dt}
    t
  \right).
  \label{eq:frequency}
\end{equation}
We note that $f(t)=\gamma H_{\rm K}m_{z}(t)/(2\pi)$ is 
the instant precession frequency of the magnetization around the easy axis. 
The initial value of the microwave frequency, $\nu(0)=c\gamma H_{\rm K}[m_{z}(0)+\epsilon]/(2\pi)$, 
has the opposite sign of the initial precession frequency of the magnetization, $f(0)=\gamma H_{\rm K} m_{z}(0)/(2\pi)$, 
under the condition of a negative $c$ 
with $\epsilon>-1(<1)$ and $\mathbf{m}(0)=+(-)\mathbf{e}_{z}$. 
Thus, for the moment, let us assume that $|\epsilon|<1$, for simplicity. 
Then, the microwaves initially rotate in the same (opposite) direction as (to) the precession 
for positive (negative) $c$, i.e., 
the sign of the parameter $c$ determines the initial direction of the microwave rotation. 
One might consider that the model becomes simple by neglecting $\epsilon$. 
As shown previously \cite{taniguchi15APEX}, however, the parameter $\epsilon$ is necessary for deterministic switching. 
Then, Eq. (\ref{eq:dEdt_initial_main}) becomes 
\begin{equation}
\begin{split}
  \frac{1}{\gamma M H_{\rm K}^{2}}
  \frac{d \mathscr{E}}{d t}
  \sim
  &
  \alpha
  \left(
    1
    -
    m_{z^{\prime}}^{2}
  \right)
  m_{z^{\prime}}
  \left[
    (c-1)
    m_{z^{\prime}}
    +
    c
    \left(
      \epsilon
      +
      \frac{d m_{z^{\prime}}}{dt}
      t
    \right)
  \right]
\\
  &+
  \frac{c}{\gamma H_{\rm K}}
  \frac{d m_{z^{\prime}}}{d t}
  m_{z^{\prime}}.
  \label{eq:dEdt_zeroth}
\end{split}
\end{equation}
The term $(dm_{z^{\prime}}/dt)m_{z^{\prime}}$ is negative 
for an excitation of the magnetization from the equilibrium state, $\mathbf{m}=\pm \mathbf{e}_{z}$, 
because $m_{z^{\prime}}$ and $dm_{z^{\prime}}/dt$ have opposite signs. 
Then, the last term, which is proportional to $c (dm_{z^{\prime}}/dt)m_{z^{\prime}}$, 
as well as $d \mathscr{E}/dt$, can be positive for a negative $c$. 
Note that Eq. (\ref{eq:dEdt_zeroth}) is valid only near the initial state. 
Therefore, this result implies the possibility of destabilizing the magnetization 
by microwaves initially rotating opposite to the precession direction. 
We also emphasize that this conclusion is obtained 
only when the microwave frequency explicitly depends on time. 
If the microwave frequency is constant, 
the last term in Eq. (\ref{eq:dEdt_zeroth}) does not appear, 
and therefore, $d \mathscr{E}/dt$ becomes negative for microwaves 
rotating opposite to the precession direction. 


The model of the phase, Eq. (\ref{eq:phase}), 
as well as the microwave frequency, Eq. (\ref{eq:frequency}), was referenced 
from our previous work corresponding to $c=1$ \cite{taniguchi15APEX}. 
The purpose of the previous work was to achieve microwave-assisted magnetization reversal 
by introducing a difference between the instant precession frequency of the magnetization, 
$\gamma H_{\rm K}m_{z}(t)/(2\pi)$, and the microwave frequency $\nu$. 
We showed in that study that the magnetization climbs up the energy landscape to synchronize the precession frequency 
with the microwave frequency and finally switches its direction to the other equilibrium. 
The parameter $\epsilon$ characterizes the difference between the instant precession frequency of the magnetization 
and the microwave frequency. 
This model originated from the recent observation of magnetization switching solely by microwaves in micromagnetic simulation \cite{kudo15}, 
as well as the experimental observation of the synchronization between the target ferromagnet and an STO \cite{suto14}. 
As pointed out in Ref. \cite{taniguchi15APEX}, 
the term proportional to $(dm_{z}/dt)t$ in Eq. (\ref{eq:frequency}) characterizing the difference between the precession frequency and the microwave frequency is 
also necessary for switching. 
The present model introduces another parameter $c$, 
which controls the sign and amplitude of the microwave frequency. 
The microwaves with positive (negative) $c$ and $|\epsilon|<1$ initially rotate 
in same (opposite) direction as (to) the magnetization precession, as mentioned above. 
In this sense, the parameter $c$ determines the rotation direction of the microwaves. 
For example, in the system where the target ferromagnet and STO are coupled, 
a positive (negative) $c$ means that 
the magnetizations in the target ferromagnet and the free layer in the STO 
initially precess in same (opposite) directions. 
It should be noted, however, that 
the microwave frequency in the present model might change its sign during precession 
because the terms $m_{z}$ and $d m_{z}/dt$ in Eq. (\ref{eq:frequency}) have opposite signs. 


Summarizing the above discussions, 
Eq. (\ref{eq:dEdt_zeroth}) implies the possibility of 
destabilizing the magnetization at an equilibrium 
by microwaves rotating opposite to the precession direction 
and having a time-dependent frequency. 
As an example, 
we consider the phase of the microwaves given by Eq. (\ref{eq:phase}). 
Two parameters, $c$ and $\epsilon$, introduced in Eq. (\ref{eq:phase}) or Eq. (\ref{eq:frequency}), 
characterize the initial rotating direction of the microwaves 
and the phase difference between the instant precession frequency of the magnetization and the microwave frequency, respectively. 
In a coupled system between a target ferromagnet and an STO, 
not only the phase $\psi$ but also the amplitude $H_{\rm ac}$ of the microwave field might depend on time. 
A way to describe such a system is to extend $c$ and/or $\epsilon$ to a complex number, $c, \epsilon \in \mathbb{C}$.



\begin{figure}
\centerline{\includegraphics[width=1.0\columnwidth]{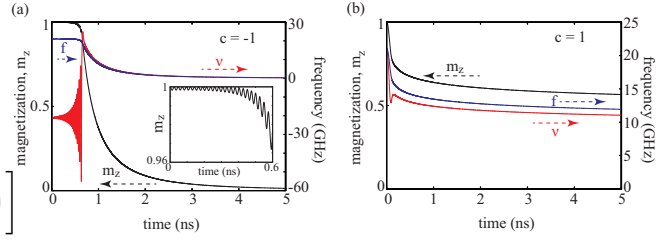}}
\caption{
         Time evolutions of $m_{z}$ (black), the instant precession frequency $f$ (blue), and the microwave frequency $\nu$ (red). 
         The inset in (a) shows the time evolution of $m_{z}$ near the initial state.
         The value of the parameter $c$ is (a) $-1$ and (b) $1$. 
         The value of the parameter $\epsilon$ is zero. 
         \vspace{-3ex}}
\label{fig:fig2}
\end{figure}



We performed numerical simulations of Eq. (\ref{eq:LLG}) to confirm the above idea. 
The values of the parameters are taken from typical experiments and numerical simulations \cite{denisov06,zhu08,okamoto08,okamoto12} as 
$M=1000$ emu/c.c., 
$H_{\rm K}=7.5$ kOe, 
$H_{\rm ac}=450$ Oe, 
$\gamma=1.764\times 10^{7}$ rad/(Oe$\cdot$s), 
and $\alpha=0.01$. 
The initial state is $\mathbf{m}(0)=+\mathbf{e}_{z}$. 
Note that the precession frequency of the magnetization, $\gamma H_{\rm K}m_{z}/(2\pi)$, 
is positive during the period from the initial state to the state where the magnetization reaches the $xy$-plane, in which $m_{z}>0$. 
Figure \ref{fig:fig2}(a) shows the time evolutions of $m_{z}$, 
the instant precession frequency $f=\gamma H_{\rm K}m_{z}/(2\pi)$, 
and the microwave frequency $\nu$ 
for $(c,\epsilon)=(-1,0)$. 
The microwave frequency, Eq. (\ref{eq:frequency}), initially takes a negative value, 
meaning that the magnetization and microwaves rotate in opposite directions. 
Nevertheless, the magnetization moves away from the initial state. 
Then, the microwave frequency rapidly changes to a positive value 
and almost becomes synchronous with the precession frequency. 
The magnetization finally arrives at the $xy$-plane, $m_{z}=0$, 
and stops its dynamics because all torques are zero on this plane. 
For comparison, the time evolutions of $m_{z}$ and $\nu$ for $(c,\epsilon)=(1,0)$ are shown in Fig. \ref{fig:fig2}(b). 
In this case, the microwaves always rotate in the same direction as the magnetization precession, i.e., $\nu>0$. 
We should emphasize here that the magnetization finally saturates at a position 
above the $xy$-plane ($m_{z} \simeq 0.6$ in this case). 
This result implies that the present model ($c=-1$) is more efficient 
of destabilizing the magnetization from equilibrium 
than our previous proposal ($c=1$) \cite{taniguchi15APEX} 
because microwaves rotating opposite to the precession direction 
do not stabilize the precession around the equilibrium. 



\begin{figure}
\centerline{\includegraphics[width=1.0\columnwidth]{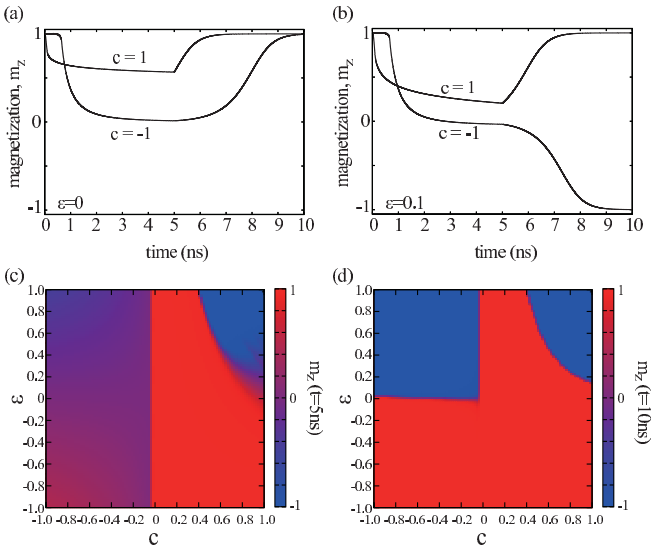}}
\caption{
         Time evolutions of $m_{z}$ for $c=\pm 1$ with (a) $\epsilon=0$ and (b) $\epsilon=0.1$. 
         The microwaves are applied from $t=0$ to $t=5$ ns 
         and are turned off at $t=5$ ns. 
         The values of $m_{z}$ at $t=5$ and $t=10$ ns are summarized in (c) and (d), respectively. 
         \vspace{-3ex}}
\label{fig:fig3}
\end{figure}



In practical application such as a magnetic recording, 
the value of $m_{z}$ after turning off the microwaves is also important. 
Therefore, we also perform numerical simulations 
in which the microwaves are applied to the ferromagnet from $t=0$ to $t=5$ ns, 
and the relaxation dynamics after turning off the microwaves are calculated from $t=5$ to $t=10$ ns. 
Figures \ref{fig:fig3}(a) and \ref{fig:fig3}(b) show examples of such dynamics, 
where the value of the parameter $\epsilon$ is (a) $0$ and (b) $0.1$. 
For $\epsilon=0$, the values of $m_{z}$ saturate above the $xy$-plane for both $c=\pm 1$. 
Therefore, after turning off the microwaves, 
the magnetization returns to the initial equilibrium state. 
When $\epsilon=0.1$, 
the magnetization moves close to the $xy$-plane, 
and for $c=-1$, the magnetization reaches below the $xy$-plane, 
and therefore, moves to the other equilibrium state after turning off the microwaves. 
On the other hand, for $c=1$, the magnetization reaches above the $xy$-plane. 
We note that this result does not contradict our previous work \cite{taniguchi15APEX}, 
in which magnetization switching was observed for $(c,\epsilon)=(1,0.1)$. 
The difference between the present and previous works is the time over which the the microwaves are applied, 
which is $5$ ns in this study while it was $50$ ns in the previous work. 
In other words, to achieve switching for $c=1$, 
the microwaves should be applied for a relatively long time 
compared with the case of $c=-1$. 
From the perspective of switching performance, 
the present model ($c<0$) will be an attractive method for fast switching. 
Figures \ref{fig:fig3}(c) and \ref{fig:fig3}(d) summarize 
the values of $m_{z}$ in the presence of the microwaves, $m_{z}(t=5{\rm \ ns})$, 
and after turning off the microwaves, $m_{z}(t=10{\rm \ ns})$. 
As shown, magnetization switching is achieved for a wide range of $\epsilon$ for $c=-1$, 
compared with the case of $c=1$. 



\begin{figure}
\centerline{\includegraphics[width=1.0\columnwidth]{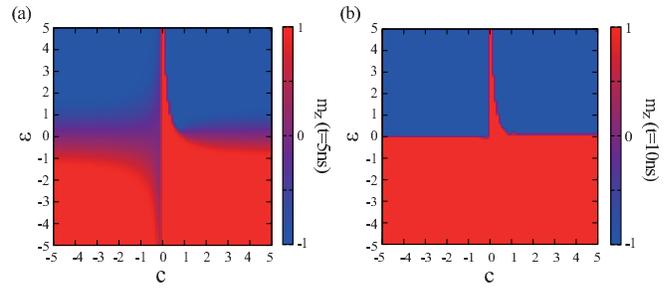}}
\caption{
         The values of $m_{z}$ for various values of $c$ and $\epsilon$ at (a) $t=5$ ns and (b) $t=10$ ns. 
         The initial state is $\mathbf{m}(0)=+\mathbf{e}_{z}$. 
         \vspace{-3ex}}
\label{fig:fig4}
\end{figure}



It is theoretically interesting to investigate the switching possibility for wider ranges of $c$ and $\epsilon$. 
In this case, however, a negative $c$ no longer guarantees that 
the microwave initially rotates opposite to the precession for $|\epsilon|>1$, as mentioned above. 
Figures \ref{fig:fig4}(a) and \ref{fig:fig4}(b) 
summarize the values of $m_{z}$ at $t=5$ and $10$ ns for wider ranges of $c$ and $\epsilon$. 
The results indicate that switching is achieved for a wide range of $c$ when $\epsilon$ is positive. 
Note that the sign of $\epsilon$ should be changed 
for switching from the opposite initial state \cite{taniguchi15APEX}. 




In conclusion, we showed that 
microwaves initially rotating opposite to the precession direction of the magnetization 
destabilize the magnetization at equilibrium 
when the microwave frequencies are time dependent. 
This argument was analytically deduced from an energy balance equation. 
We proposed a model of the system having a time-dependent microwave frequency 
with two parameters, $c$ and $\epsilon$ 
to test the possibility of switching by such microwaves. 
The parameters $c$ and $\epsilon$ characterize the rotation direction of the microwaves near the initial state 
and the difference between the instant precession frequency of the magnetization and the microwave frequency, respectively. 
We confirmed magnetization switching solely by microwaves 
rotating initially opposite to the precession direction by numerical simulation. 
We also showed that the use of such microwaves has the possibility of realizing fast switching 
compared with microwaves always rotating in the same direction.


The author expresses gratitude to 
Shinji Yuasa, Kay Yakushiji, Hitoshi Kubota, Akio Fukushima, Takehiko Yorozu, 
Yoichi Shiota, Sumito Tsunegi, Satoshi Iba, 
Aurelie Spiesser, Hiroki Maehara, 
and Ai Emura for their support and encouragement. 
This work was supported by JSPS KAKENHI Grant-in-Aid for Young Scientists (B) 25790044. 




\end{document}